\documentclass[12pt]{article}
\usepackage{amssymb,amsmath,epsfig}
\allowdisplaybreaks

\begin{document}
\title{\bf Gravastars in $f(R,T,R_{\mu\nu}T^{\mu\nu})$ Gravity}
\author{Z. Yousaf \thanks{zeeshan.math@pu.edu.pk}, M. Z. Bhatti
\thanks{mzaeem.math@pu.edu.pk} and H. Asad \thanks{hamnaasad96@gmail.com}\\
Department of Mathematics, University of the Punjab,\\
Quaid-i-Azam Campus, Lahore-54590, Pakistan.}

\date{}

\maketitle
\begin{abstract}
This work is devoted to study the analytical and regular solutions
of a particular self-gravitating object (i.e., gravastar) in a
particular theory of gravity. We derive the corresponding field
equations in the presence of effective energy momentum tensor
associated with the perfect fluid configuration of a spherical
system. We then describe the mathematical formulations of the three
respective regions i.e., inner, shell and exterior of a gravastar
separately. Additionally, the significance and physical
characteristics along with the graphical representation of
gravastars are discussed in detail. It is seen that under some
specific constraints, $f(R,T,R_{\mu\nu}T^{\mu\nu})$ gravity
is likely to host gravastars.
\end{abstract}


\section{Introduction}

General relativity (GR) laid down the foundation of new era of
mathematical physics that has led to the discovery of the new ideas
in the field of astrophysics, cosmology, etc. Einstein came up with
the notion of matter-geometry relationship that led him to explain
the concept of gravity in terms of curvature in spacetime. There are some surprising and revolutionary results based on some
notable observations, like cosmic microwave background radiation,
type Ia supernovae, etc., \cite{1}-\cite{3} suggested the accelerated
expansion of our universe. Modified theories of gravity are among
the most popular approaches considered for the explanation of the
structure and origins of the universe. Furthermore, the
modification of GR may lead us to explain the history and the
accelerated expansion of the universe, and the role of dark matter
in a better way. Nojiri and Odinstov \cite{4}
reformulated the Hilbert action by considering the generic function
of Ricci scalar i.e., $f(R)$ in place of $R$, thereby opening a arena
for modified theories of gravity. Afterwards, many facts and
unsorted problems about the structure of different stellar objects
and mystery of universe were discussed \cite{5}-\cite{17}.

The generalization of $f(R)$ was unveiled by Harko \emph{et al.} \cite{18} named as $f(R,T)$ in where $T$ is
the trace of energy momentum tensor. The motivation of including $T$ in $f(R)$ is that it may give us viable results with respect to quantum gravity by retaining coupling between matter and geometry.
They considered some definite forms of the function
$f(R,T)$ and evaluated the corresponding field as well as conservation equations.
Moreover, they examined the scalar field model in detail. In order to
do so, they acknowledged the cosmological significance of
$f(R,T^{\phi})$ model. Furthermore, they demonstrated the equation of
motion of test particle and the Newtonian limit of equation of
motion.

Likewise, to acquire information on how the strong non-minimal
coupling of matter and geometry influenced the results of $f(R,T)$
theory, the generalization in this theory has been developed.
Subsequently, Haghani \emph{et al.} \cite{19}
introduced the $f(R,T,Q)$ theory ($Q\equiv R_{\mu\nu}T^{\mu\nu}$) which involves the contraction of
the Ricci scalar with the stress energy momentum tensor. They
deduced the field equations for the motion of test particle in the
$f(R,T,Q)$ gravity for both the conservative and
non-conservative cases. They claimed that this theory is non-conserved
because of the extra force acting on the particles. The
Dolgov-Kawasaki instability is also examined along with the stability
condition. In addition to this, some cosmological applications were
studied, analytically and numerically by Ayuso \emph{et al.} \cite{20}.
Odintsov and S\'{a}ez-G\'{o}mez \cite{21} demonstrated
the phenomenology of the $f(R,T,Q)$ gravity and
analyzed the instability in this gravity. Baffou \emph{et al.}
\cite{22} explored the model in
$f(R,T,Q)$ gravity to discuss the stability
utilizing the de-Sitter and power law solutions. They also evaluated
the generalized Friedmann equations. The $f(R,T,Q)$ gravity enforced
to look upon the universe extensively. Dvali
\cite{23} explained thoroughly some of the
properties of the theories that can reformulate gravity.
Yousaf \emph{et al.} \cite{24}-\cite{26} studied the phenomena of dynamical instability
in $f(R,T,Q)$ gravity by taking into account different interior geometries. They used a specific Harrison-Wheeler equation of state in order to describe stable regions of the corresponding stellar structures. Furthermore, Yousaf \emph{et al.} \cite{27} calculated junction condition in order to match exterior Einstein-Rosen bridge with an interior cylindrical spacetime.

Existence of a new hypothetical star can be evident as a consequence
of a gravitational collapse. It can be attained by scrutinizing the
basic concept of Bose-Einstein condensation (BEC). In BEC, the gas of bosons are cooled to the temperature where the
molecules loses infinite kinetic energy. One can recognize that all the
molecules of bosons are indistinguishable having
same quantum spin, thereby not following Pauli exclusive
principle. Here, the molecules continue to move slowly so that they
must obey the uncertainty principle. The one of the final states of gravitational collapse could give rise to gravastar, thereby indicating this
as an alternative to black hole as suggested by \cite{28}. Such kind of structure is mainly
comprised of three regions, namely, inner region, shell and
exterior region. The inner region is supplemented by that matter in which the energy
density $\rho$ is equal to the negative pressure
$P$ i.e., $\rho=-P$. This state is the main hinderance
that doed not permit gravastar to form singularity. The presence of
dark energy in the inner region is a source of repulsive force that
is exerting on the shell. The shell contains the effective matter
incorporate with the perfect fluid. In this
region, the energy density $\rho$ is equal to the
 pressure $P$ of the relativistic fluid, i.e., $\rho=P$. The
shell in return apply force on the inner region so that the
gravastar would remain in the hydrostatic equilibrium. The exterior region
is entirely vacuum. It is an indication of the negligible
atmospheric pressure out there. So the EoS $\rho=P=0$ is adopted for
this region. One can see the formation of naked singularity and black holes phenomenon with the help of mathematical
equation presented by Virbhadra \emph{et al.}
\cite{29}, \cite{30}. Gravitational collapse as well as the stability of
self-gravitating systems have also been analyzed in literature \cite{31}-\cite{34}.

Mazur and Mottola \cite{28} presented the visualization of the gravastar and claimed the
absence of event horizon and singularity in such structures. They
also discussed the thermodynamic stability of the gravastar. Later
Visser and David \cite{35} extended the work of Mazur
and Mottola and examined the stability of this star by taking into
account the physical characteristics and equation of state. Cattoen \emph{et al.} \cite{36} used the concept
that if there does not exist any shell in the composition of
gravastar then it must have anisotropic pressure. Chirenti
and Rezzolla \cite{37} demonstrated the stability of a
gravastar against the generic perturbation scheme. Additionally, gravastar
can be distinguished from the black hole of same mass as their
quasi-normal modes are different. Bili\'{c} \emph{et al.}
\cite{38} formulated some solutions of gravastar
through Born infeld Phantom background. Horvat \emph{et al.} \cite{39} performed stability analysis for the static spherically symmetric gravastars and observed that anisotropic gravastar structures exhibiting continuous pressure are radially stable.

Sakai \emph{et al.} \cite{40} examined the optical images of gravastar
composed of unstable circular orbit. In fact, they assumed two optical
sources and described both of them in detail. Chirenti \emph{et al.}
\cite{41} worked on the ergoregion instability
in a rotating gravastar and declared it to be unstable. Rahaman \emph{et al.} \cite{42} studied the role of electric charge on gravastar with (2+1)-dimensional metric. They studied some theoretical results for the stability of spherical gravastars. Pani \emph{et al.} \cite{43} constructed non-rotating gravastar models after matching smoothly Schwarzschild and de Sitter
spacetimes.

Ghosh \emph{et al.} \cite{44} analyzed the existence of these structures
by taking higher dimensional geometries. Das \emph{et
al.} \cite{45} extended the concept of gravastars $f(R,T)$ gravity and checked some important characteristics about
their theoretical existence. Shamir and Mushtaq \cite{46} modified their analysis for $f(G,T)$ gravity. Yousaf \cite{47,48} found some mathematical models of compact objects in $f(G,T)$ gravity through structure scalars.
Yousaf \emph{et al.} \cite{49} analyzed the effects of electromagnetic field on the possible formation of gravastars in $f(R,T)$
gravity. Bhatti and his collaborators \cite{50}, \cite{51} extended these results for spherically and cylindrically symmetric spacetime in GR. Sharif and Waseem \cite{52} described the modeling of gravastars with the help of conformal motion. Recently, Yousaf \cite{53} found non-singular solutions of gravastars in $f(R,T)$ gravity by taking static cylindrically symmetric spacetime. He found the existence of cylindrical gravastar-like structures in $f(R,T)$ gravity.

The present work is devoted to study the
formation of gravastars for the isotropic self-gravitating spherical
system in $f(R,T,Q)$ theory. In the next section the field as well
as the basic equation for $f(R,T,Q)$ theory have been derived. The
construction of gravastars including the detailed discussion about
the modeling of all the three regions of gravastars are discussed in
section 3. In section 4, the appropriate junction conditions are
calculated, while the section 5 is devoted to describe various
physical characteristics  of the gravastar in $f(R,T,Q)$ theory along
with their graphical representation.
Finally in section 6, we sum up all the results.

\section{$f(R,T,Q)$ Theory of Gravity}

The motivation to adopt $f(R,T,Q)$ theory is that in these type of theories the more generalized form of the matter Lagrangian $L_{m}$ has been used describing the strong non-minimal coupling between the matter and geometry contrary to the $f(R,T)$ theory. For instance, $f(R,T,Q)$ can be generalized by the addition of the term $Q\equiv R_{\mu\nu}T^{\mu\nu}$ in the Lagrangian. Einstein Born Infeld theories are the example of such coupling. The thought-provoking distinction between the $f(R,T,Q)$ and $f(R,T)$ theories is that even if we consider the traceless energy momentum tensor i.e., $T=0$ in $f(R,T,Q)$, this theory will be able to explain about the non-minimal coupling to the electromagnetic field due to the addition of term $R_{\mu\nu}T^{\mu\nu}$ in the Lagrangian.

The $f(R,T,Q)$ theories can explain the late time acceleration without using the approach of the cosmological constant or dark energy. In contrast with the $f(R,T)$ theories, these theories involve the effects of the extra force exerted on the massive particle even if we consider $L_{m}=-\rho$, where $\rho$ is the energy density of the relativistic fluid. The characteristics of galactic rotation curves could be explained in the presence of this extra force without considering the hypothesis of dark matter. Hence, the complex modified equations are attained in this theory. One can achieve the various qualitative cosmological solutions with different forms of function in the $f(R,T,Q)$ theories. This theory makes feasible for us to reveal about the evolution in the universe at the early stages instead of using the inflationary paradigm, which is quite challenging.

Odintsov and S{\'a}ez-G{\'o}mez \cite{21} demonstrated that the finest power law version of $f(R,T,Q)$ theories might be represented by the Horava-like gravity power-counting renormalizable covariant gravity. The action integral for the case of $f(R,T,Q)$ theory can be stated as
\begin{equation}\label{1}
S=\frac{1}{2}\left(\int d^{4}xf(R,T,R_{\alpha\beta}T^{\alpha\beta})\sqrt{-g}+\int d^{4}xL_{m}\sqrt{-g}\right),
\end{equation}
where $R$ is the Ricci scalar, $Q\equiv R_{\alpha\beta}T^{\alpha\beta}$, $L_m$ is the matter Lagrangian, $T$ and $g$ are the traces of energy momentum and metric tensors, respectively. By varying the
above action respecting $g_{\mu\nu}$, one can get
\begin{eqnarray}\nonumber
&-&G_{\alpha\beta}(f_{Q}L_{m}-f_{R})-g_{\alpha\beta}\left\{\frac{f}{2}-\Box
f_{R}-\frac{R}{2}f_{R}-\frac{1}{2}\nabla_{\pi}\nabla_{\rho}(f_{Q}T^{\pi\rho})-L_{m}f_{T}\right\}\\\nonumber
&+&2f_{Q}R_{\pi(\alpha}T_{\beta)}^{\pi}+\frac{1}{2}\Box(f_{Q}T_{\alpha\beta})-\nabla_{\pi}
\nabla_{(\alpha}[T_{\beta)}f_{Q}]-2(f_{T}g^{\pi\rho}+f_{Q}R^{\pi\rho})\frac{\partial^{2}L_{m}}{\partial
g^{\alpha\beta}\partial g^{\pi\rho}}\\\label{q2}
&-&T_{\alpha\beta}^{(m)}(f_{T}+\frac{R}{2}f_{Q}+1)-\nabla_{\alpha}\nabla_{\beta}f_{R}=0,
\end{eqnarray}
where $\nabla_\alpha$ is a covariant derivative,
$\Box=g^{\alpha\beta}\nabla_{\alpha}\nabla_{\beta}$, $G_{\lambda\sigma}$ is an Einstein tensor and
$T_{\lambda\sigma}$ describes energy-momentum tensor that can be
described as follows
\begin{align}\label{q3}
&T_{\lambda\sigma}^{(m)}=-\frac{2}{\sqrt{-g}}\frac{\delta(\sqrt{-g}\textit{L}_m)}
{\delta{g^{\lambda\sigma}}}.
\end{align}
Here, the subscripts $R,~T$ and $Q$ indicate the partial derivative of $f$ with respect to its arguments. From Eq.(\ref{q3}), one can write
\begin{align}\nonumber
&3\Box
f_R+\frac{1}{2}\Box(f_QT)-T(f_T+1)+\nabla_\pi\nabla_\rho(f_QT^{\pi\rho})+
R(f_R-\frac{T}{2}f_Q)\\\nonumber &+(Rf_Q+4f_T)\textit{L}_m
-2f+2R_{\pi\rho}T^{\pi\rho}f_Q -2
\frac{\partial^2\textit{L}_m}{\partial g^{\lambda\sigma}\partial
g^{\pi\rho}}\left(f_Tg^{\pi\rho}+f_QR^{\pi\rho}\right).
\end{align}
By applying the conditions of vacuum case on the field equation of
$f(R,T,Q)$ theory, equations of motion for $f(R)$ theory can be
retrieved. Nonetheless, the case $Q=0$ provides the dynamics of
$f(R,T)$ theory. It is important to stress that the value of the
last term will be non-zero, if $\textit{L}_m$ contains second or
higher order terms. However, by setting a particular choice of
$L_m$, one can ignore this last term from the calculations. The
$f(R,T,Q)$ theory is an un-conserved theory which eventually leads
an extra force (even within our assumed case $L_{m}=-\rho$). Our present theory
is a special form of gravitational theory. The non-minimal coupling
between matter and geometry can be dissolved in $f(R,T)$ gravity,
when one consider a charged gravitating sources in the analysis.
This would boils down $f(R,T)$ equation of motion to that of $f(R)$.
However, in this theory, the term $Q$ preserves such non-minimal
coupling of gravity with the relativistic fluid. One can write
Eq.\eqref{q2} in the following alternative form as
\begin{equation}\label{2}
G_{\nu}^{\mu}=T_{\nu}^{\mu(eff)},
\end{equation}
where
\begin{eqnarray}\nonumber
 T_{\mu\nu}^{(eff)}&=&\frac{1}{f_{R}-L_{m}f_{Q}}\left[\left(f_{T}+\frac{1}{2}Rf_{Q}+1\right)
 T_{\mu\nu}^{(m)}+\left\{\right.\frac{f}{2}-\frac{f_{R}R}{2}-L_{m}f_{T}\right.
\\\nonumber &-&\frac{1}{2}\nabla_{\alpha}\nabla_{\beta}(f_{Q}T^{\alpha\beta})\left.\right\}g_{\mu\nu}
-\frac{1}{2}\Box(f_{Q}T_{\mu\nu})
-(g_{\mu\nu}g^{\alpha\beta}\nabla_{\alpha}\nabla_{\beta}-\nabla_{\mu}\nabla\nu)f_{R}\\\label{2a}
&-&2f_{Q}R_{\alpha(\mu}T_{\nu)}^{\alpha}
+\nabla_{\alpha}\nabla_{(\mu}[T_{\nu)}^{\alpha}f_{Q}\left.\right]
+\left.2(f_{Q}R^{\alpha\beta}+f_{T}g^{\alpha\beta})\frac{\partial^2 L_{m}}{\partial g^{\mu\nu}\partial g^{\alpha\beta}}\right],
\end{eqnarray}
could be regarded as an effective energy-momentum tensor for
$f(R,T,Q)$ gravity. The purpose of the present paper is to analyze
the effects of $f(R,T,Q)$ terms on the stability of gravastars. For
this purpose, we consider an isotropic matter distribution having
the following form
\begin{eqnarray}\label{4}
T_{\mu\nu}^{(m)}&=(\rho+P)u_{\mu}u_{\nu}-Pg_{\mu\nu},
\end{eqnarray}
where $u^\mu$ is the four velocity of the fluid and $P$ stands for the fluid's pressure.

\section{Formation of Static Spherically Symmetric Gravastars in $f(R,T,Q)$ Gravity}

Now, we must assume a static irrotational spherically symmetric spacetime, which in Schwarzschild-like coordinate system, can be expressed as
\begin{equation}\label{7}
ds^{2}=e^{\nu(r)}dt^{2}-e^{\mu(r)}dr^{2}-r^{2}(d\theta^{2}+\sin^{2}\theta d\phi^{2}).
\end{equation}
The non-zero components of Einstein tensor for the spherically symmetric spacetime are found as follows
\begin{eqnarray}\label{8}
  G_{0}^{0}&=&\frac{e^{-\lambda}}{r^{2}}(-1+e^{\lambda}+\lambda'r),\\\label{9}
  G_{1}^{1}&=&\frac{e^{-\lambda}}{r^{2}}(-1+e^{\lambda}-\nu'r),\\\label{10}
  G_{2}^{2}=G_{3}^{3}&=&\frac{e^{-\lambda }}{4r} \left[2(\lambda'-\nu')-(2\nu''+\nu'^{2}-\nu'\lambda')r\right],
\end{eqnarray}
where prime denotes derivatives with respect to $r$. The modified field equations can be found after using Eqs.\eqref{2}, \eqref{2a}, \eqref{4}, \eqref{7}, \eqref{8}, \eqref{9} and \eqref{10} as
\begin{align}\nonumber
&\frac{e^{-\lambda}}{r^{2}}\left(-1+e^{\lambda}+\lambda'r\right)=\frac{1}{f_{R}+\rho
f_{Q}}\left[\right.\rho\chi_{1}+\rho'\chi_{2}+\frac{\rho''f_{Q}e^{-\lambda}}{2}+P\chi_{3}\\\label{11}
&+P'\chi_{4}-\frac{P''f_{Q}e^{-\lambda}}{2}+D_{0}\left.\right],\\\label{12}
&\frac{e^{-\lambda}}{r^{2}}\left(-1+e^{\lambda}-\nu'r\right)=\frac{-1}{f_{R}+\rho
f_{Q}}\left[\rho\chi_{5}-\frac{\rho'\nu'e^{-\lambda}f_{Q}}{4}+P\chi_{6}+P'\chi_{7}+D_{1}\right],\\\nonumber
&\frac{e^{-\lambda}}{4r}\left[2(\lambda'-\nu')-(2\nu''+\nu'^{2}-\nu'\lambda')r\right]=\frac{-1}{f_{R}+\rho
f_{Q}}\left[\right.\chi_{8}
+\frac{\rho'\nu'f_{Q}e^{-\lambda}}{4r^{2}}+P\chi_{9}\\\label{13}
&+P'\chi_{10}
  +P''f_{Q}e^{-\lambda}+D_{2}\left.\right].
\end{align}
The values of $\chi_{i}'s$ and that of $D_{i}'s$ are given in the Appendix.

The divergence of effective energy momentum
tensor in this modified theory gives rise to
\begin{align}\label{6}
\nabla^\lambda
T_{\lambda\sigma}&=\frac{2}{Rf_Q+2f_T+2}\left[\nabla_\sigma(\textit{L}_mf_T)
+\nabla_\sigma(f_QR^{\pi\lambda}T_{\pi\sigma})-\frac{1}{2}(f_Tg_{\pi\rho}+f_QR_{\pi\rho})\right.\\\nonumber
&\times\left.\nabla_\sigma
T^{\pi\rho}-G_{\lambda\sigma}\nabla^\lambda(f_Q\textit{L}_m)\right],
\end{align}
which for our observed system, i.e., \eqref{4} and \eqref{7} provide
\begin{equation}\label{14}
\frac{\partial P^{eff}}{\partial r}+\frac{\nu'}{2}\left[\rho^{eff}+P^{eff}\right]-Z=0,
\end{equation}
where the term $Z$  appears due to non-conserved nature of this
theory. This term would allow non-geodesic motion of the test
particles and is given in Appendix. One can write from Eq.\eqref{11}
the value of the metric coefficient with respect to the mass of the
spherical system ($m$) in the presence of extra curvature terms as
under
\begin{equation}\label{15}
e^{-\lambda}=1-\frac{2m}{r}+N(r)\\,
\end{equation}
where,
\begin{eqnarray}
N(r)=\int\rho^{*(eff)}r^{2}dr.
\end{eqnarray}
Now, we would like to evaluate above equation by taking into account
constant $R,~T$ and $Q$ values. In this background, the metric
coefficient \eqref{15} becomes
\begin{equation}\label{n15}
e^{-\lambda}=1-\frac{2m}{r}+\frac{\rho^{*(eff)}r^{3}}{3}.
\end{equation}
To obtain hydrostatic equilibrium in the $f(R,T,Q)$ theory, we use Eqs.\eqref{12}, \eqref{14} and \eqref{15} to get
\begin{eqnarray}\nonumber
  \frac{\partial P^{(eff)}}{\partial r} &=&\frac{-2r^{2}(\rho\chi_{5}+P\chi_{6}+P'\chi_{7}+D_{1})
  +H(\frac{2m}{r}+N(r))}{1-\frac{2m}{r}+N(r)(2rH)}[\rho^{(eff)}\\\label{16}
  &+&P^{(eff)}]+Z.
\end{eqnarray}
Under constant curvature condition, the above equation reduces to
\begin{eqnarray}\nonumber
 \frac{\partial P^{(eff)}}{\partial r} &=&\frac{-2r^{2}(\rho\chi_{5}+P\chi_{6}+P'\chi_{7}+D_{1})+H(\frac{2m}{r}+\frac{\rho^{*(eff)}r^{3}}{3})}
 {1-\frac{2m}{r}+\frac{\rho^{*(eff)}r^{3}}{3}(2rH)} [\rho^{(eff)}\\\label{n16}
 &+&P^{(eff)}]+Z.
\end{eqnarray}
In the following subsections, we shall describe the construction of
isotropic spherical gravastar structures. For this purpose, we shall
apply the EoS of the corresponding three regions on the mathematical
formulations in order to get the desired results. We shall also take
exterior spacetime in order to match them on the hypersurface with
an appropriate interior one. This study will be accompanied by the
analysis of ultrarelativistic fluid within the shell.

\subsection{Region I}

In order to explain region I, we shall consider an EoS that can be expressed in terms of effective $f(R,T,Q)$ matter variables as
\begin{equation}
P=-\rho.
\end{equation}
One can see that this EoS is a  barotropic EoS, i.e.,
$P=\omega\rho$. The selection of $\omega=-1$ in the
said EoS acts as the cosmological constant ($\Lambda$). Thus the EoS
with $\omega=-1$ could be considered as the best and simplest model
to describe dark energy among all the proposed dark energy models
available in the literature. Through experimental data, observed
value of $\Lambda$ has been found to be $10^{47}GeV^4$, in order
$\Lambda$ to describe as a vacuum energy. So, we have
\begin{equation}\label{18}
\rho=\rho_{0}(constant),
\end{equation}
which eventually gives
\begin{equation}\label{19}
P=-\rho_{0}.
\end{equation}
Putting Eq.\eqref{19} in Eq.\eqref{11}, we have the metric potential
$\lambda$ as follows
\begin{equation}\label{21}
e^{-\lambda}=1-\frac{1}{r}\int\frac{r^{2}}{H}[\rho_{0}(\chi_{1}-\chi_{3})+D_{0}]dr,
\end{equation}
where we have neglected the value of an integration constant. This
has assumed to be zero, in order to have a regular solution around the
center of the gravastar. Further, in the above equation, the
quantity $D_{0}$ describes the portion of the equation containing
$f(R,T,Q)$ gravity terms. With constant $R,~T$ and $Q$ condition,
Eq.\eqref{21} turns out to be
\begin{equation}\label{21n}
e^{-\lambda}=1-\frac{r^{2}}{3H}\left[\rho_{0}(\chi_{1}-\chi_{3})+D_{0}\right].
\end{equation}
After using Eqs.\eqref{11}, \eqref{12}, \eqref{18} and \eqref{19},
one can write a correlation between two unknowns $\nu$ and $\lambda$
as follows
\begin{equation}\label{22}
e^{\nu}=e^{-\lambda}+V(r),
\end{equation}
where
\begin{eqnarray*}
 V(r)=\int\frac{1}{rH}[\rho_{0}(\chi_{1}-\chi_{3}+\chi_{5}-\chi_{6})+D_{0}+D_{1}]dr.
\end{eqnarray*}
With constant $R,~T$ and $Q$ condition, Eq.\eqref{22} becomes
\begin{equation}\label{22n}
e^{\nu}=e^{-\lambda}+r^{\frac{1}{H}[\rho_{0}(\chi_{1}-\chi_{3}+\chi_{5}-\chi_{6})+D_{0}+D_{1}]}+B.
\end{equation}
The gravitational mass is calculated as:
\begin{equation}
M(D)=\int_{0}^{D}\frac{\rho_{0}r^{2}}{2}dr=\frac{D^{3}}{6}\rho_{0}^{(eff)}.
\end{equation}
This equation clearly shows a definite connection between both the
gravitating matter source of the self-gravitating system and its
radial coordinates, which can be taken to be one of the very
significant characteristics of relativistic compact bodies. This
also points out the direct dependance of the mass $M$ with a
specific radial distance.

\subsection{Region II}

The region II corresponds to the shell of the gravastars. We model
the shell to consists of ultrarelativistic fluid obeying a specific
configurations of EoS. We propose that within the shell, the
 energy density is equal to the  pressure, i.e.,
$\rho=P$. It has the thick layer but that thickness
is very small, as it is $e^{-\lambda}$ and the value of $\lambda$ is
less than 1, thereby making the shell's thickness in between the 0
to 1. These suppositions also makes our calculations easy for us to
deal with them. Here we reviewed the form of fluid conceived by
Zel'dovich \cite{54}. Such kind of fluid has
been considered by most of the cosmological
\cite{55} and astrophysical
\cite{56,57,58} fields. After
solving simultaneously the three non-zero field equations along with
some viable assumptions described in \cite{53}, we get the
following two equations as
\begin{align}\label{24}
&\frac{d}{dr}(e^{-\lambda})=\frac{2}{r}+\int\frac{r\psi_{1}}{H}dr,\\\label{25}
&\left[\frac{\nu'}{4}-\frac{1}{2r}\right]\frac{d}{dr}(e^{-\lambda})=\psi_{2}-\frac{1}{r^{2}},
\end{align}
where
\begin{eqnarray*}
\psi_{1}&=&\rho(\chi_{1}+\chi_{3}+\chi_{5}+\chi_{6})+\rho'(\chi_{2}+\chi_{4}+\chi_{7}-\frac{\nu'e^{-\lambda}f_{Q}}{4})+D_{0}+D_{1},\\\nonumber
\psi_{2}&=&\frac{-1}{H}\left[\rho(\chi_{5}+\chi_{6}+\chi_{8}+\chi_{9})+\rho'(\chi_{7}+\chi_{10}+\frac{\nu'f_{Q}e^{-\lambda}}{4r^{2}}-\frac{\nu'f_{Q}e^{-\lambda}}{4})\right]\\\nonumber
&+&\rho''f_{Q}e^{-\lambda}+D_{1}+D_{2}.
\end{eqnarray*}
where $\psi_{1}$ contains the influence of effective matter with the
involvement of dark source terms. The solution of Eq.\eqref{24}
provides
\begin{equation}\label{26}
e^{-\lambda}=2\ln[r]+\int\left(\int\frac{r\psi_{1}}{H}dr\right)dr+C.
\end{equation}
where $C$ is the constant of integration. It is worthy to describe
that Eq.\eqref{26} gives us the value of thickness $(e^{-\lambda})$
of shell. The radius $r$ varies from $D$ to $D+\epsilon$. If we have
$\epsilon<1$ in that case we have $C<1$ and hence
$e^{-\lambda}\ll1$. Evaluating Eqs.\eqref{24} and \eqref{25}, we
obtain
\begin{equation}\label{27}
e^{\nu}=e^{2\psi_{3}}+F,
\end{equation}
where
\begin{eqnarray}\nonumber
&&\psi_{3}=\int\frac{2r\psi_{2}(r)+\psi_{2}^{*}(r)}{2+r\psi_{2}^{*}(r)}dr,\quad
\psi_{2}^{*}(r)=\int\frac{r\psi_{1}}{H}dr.
\end{eqnarray}
Now, after considering EoS for the thin shell in
Eq.\eqref{14}, we get
\begin{equation}\label{28}
\rho^{(eff)}=P^{(eff)}=\frac{Z}{\nu'}-\int\frac{Z'}{\nu'}dr,
\end{equation}
where
\begin{align}\nonumber
Z&=\frac{2}{2+Rf_{Q}+2f_{T}}\left[\right.\rho'f_{Q}e^{-\lambda}\left(\frac{5\nu'\lambda'}{8}-\frac{5\nu'^{2}}{8}-\frac{5\nu''}{4}-\frac{5\nu'}{2r}\right)
-\frac{\rho'f_{T}e^{\nu}}{r^{2}}\\\nonumber
&+\rho f_{Q}e^{-\lambda}
\left(\frac{\nu'\lambda'^{2}}{4}-\frac{\nu''\lambda'}{2}-\frac{\lambda'\nu'^{2}}{4}+\frac{\lambda'^{2}}{r}+\frac{\nu''}{r}
-\frac{\nu'\lambda'}{r}-\frac{\nu'}{r^{2}}-\frac{2}{r^{3}}+\frac{2e^{\lambda}}{r^{3}}\right)\left.\right].
\end{align}

\subsubsection{Constant $R,~T$ and $Q$}

\begin{center}\begin{figure} \centering
\epsfig{file=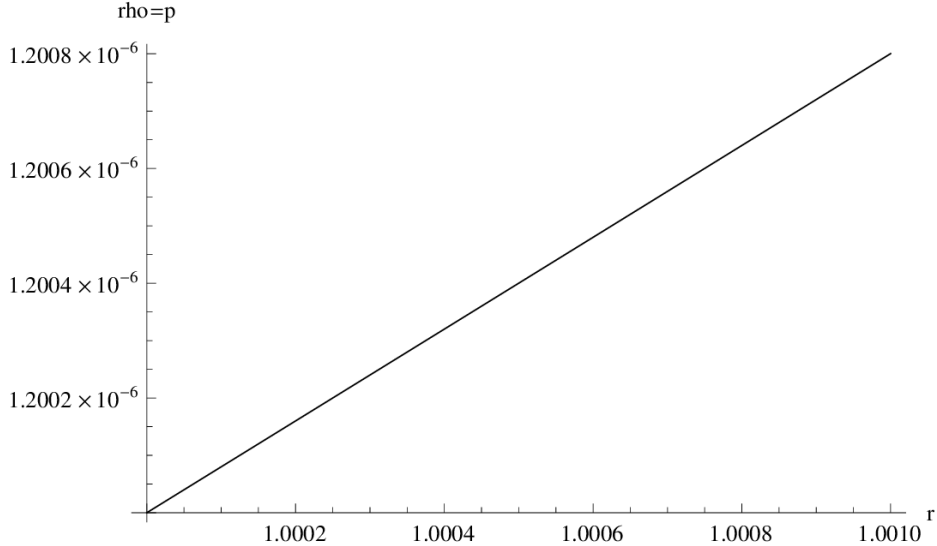,width=0.9\linewidth} \caption{Variations in
the profile of density relationship with respect to $r$} \label{f1}
\end{figure}\end{center}

In this subsection, we shall take constant values of curvature
terms. In this context, Eq.\eqref{24} turns out to be
\begin{align}\label{24n}
&\frac{d}{dr}(e^{-\lambda})=\frac{2}{r}+\frac{r\psi_{1}}{H},
\end{align}
whose integration provides
\begin{equation}\label{26n}
 e^{-\lambda}=2\ln[r]+\frac{r^{2}\psi_{1}}{2H}+C,
\end{equation}
where $C$ is an integration constant. Furthermore, an equation
analogous to Eq.\eqref{27} is found to be
\begin{equation}
e^{\nu}=F(2+r^{2}\psi_{1})^{\frac{\psi_{3}}{2\psi_{1}}},
\end{equation}
where
\begin{eqnarray*}
\psi_{3}=4H\psi_{2}+2\psi_{1}.
\end{eqnarray*}
By making use of EoS $\rho=P$ along with
Eq.\eqref{14} with constant curvature constraints, we get an
equation which states that effective energy density of the shell is
more dense than that of the inner region. The figure \eqref{f1}
verified our supposed equation of state i.e., $\rho=P$
and shows direct relationship between them.

\subsection{Region III}

As discussed earlier, it is seen that in the region III, one needs
to take $\rho=P=0$ . In this framework, we consider
a vacuum exterior region that can be illustrated through a
Schwarzschild solution as follows
\begin{equation}
ds^{2}_+=\left(1-\frac{2M}{r}\right)dt^{2}-\left(1-\frac{2M}{r}\right)^{-1}dr^{2}-r^{2}(d\theta^{2}+\sin^{2}\theta
d\phi^{2}),
\end{equation}
where $M$ is the total gravitating mass of the system.

\section{Junction Conditions}

This section is devoted to find the some suitable constraints that
can be helpful for the smooth joining of exterior and interior
metrics. One can get the required condition with the help of
Darmois-Israel formalism \cite{59,60}. In this formalism, the metric coefficients
must be continuous at the junction of two regions at $r=D$ but their
derivatives may not be necessarily continuous. Further, working on
this formalism we can successfully calculate the stress-energy
tensor $S^{i}_{j}$. The Lanczos equation
\cite{61,62,63,64}
could play a significant role in determining the intrinsic surface
stress-energy tensor as follows
\begin{equation}\label{30}
S^{i}_{j}=-\frac{1}{8\pi}(\kappa^{i}_{j}-\delta^{i}_{j}\kappa^{k}_{k}),
\end{equation}
where $\kappa^{i}_{j}=K^{+}_{ij}-K^{-}_{ij}$ which indicates that
how much extrinsic curvatures are discontinuous over the boundary
surface. Here $-$ sign indicates the interior region and the $+$
shows the exterior region. The second fundamental forms linked with
both sides of the shell can be stated as
\begin{equation}\label{31}
K^{^{+}_{-}}_{ij}=-n^{^{+}_{-}}_{\nu}\left[\frac{\partial^{2}x_{\nu}}{\partial\xi^{i}\partial\xi^{j}}+\Gamma_{\alpha\beta}^{\nu}\frac{\partial
x^{\alpha}}{\partial\xi^{i}}\frac{\partial
x^{\beta}}{\partial\xi^{j}}\right]|_{\Sigma},
\end{equation}
where $\xi^{i}$ represents the intrinsic coordinates on the shell,
$n_{\nu}^{^{+}_{-}}$ denotes the unit normal to the surface
$\Sigma$. The induced spherically symmetric static metric on the
hypersurface takes the form
\begin{equation}\label{32}
ds^{2}=f(r)dt^{2}-\frac{dr^{2}}{f(r)}-r^{2}(d\theta^{2}+\sin^{2}\theta
d\phi^{2}),
\end{equation}
and for this type of metric $n_{\nu}^{^{+}_{-}}$ can be described as
\begin{equation}\label{33}
n^{^{+}_{-}}_{\nu}=^{+}_{-}|g^{\alpha\beta}\frac{\partial
f}{\partial x^{\alpha}}\frac{\partial f}{\partial
x^{\beta}}|^{\frac{-1}{2}}\frac{\partial f}{\partial x^{\nu}},
\end{equation}
with $n^{\mu}n_{\mu}=1$.

Taking into account the Lanczos equation the stress energy tensor
$S^{i}_{j}=diag[\sigma,-\nu,-\nu,-\nu]$ can be easily evaluated.
Here, the structural variables, energy density and pressure on the
surface are denoted with $\sigma$ and $\nu$, respectively. The
generic formula for the computation of surface energy density and
the surface pressure can be formulated as under
\begin{align}\label{34}
&\sigma=\frac{-1}{4\pi D}[\sqrt{f}]^{+}_{-},\\\label{35}
&\nu=\frac{-\sigma}{2}+\frac{1}{16\pi}\left[\frac{f'}{\sqrt{f}}\right]^{+}_{-},
\end{align}
By working on Eqs.\eqref{34} and \eqref{35}, we come up with
\begin{align}\label{36}
&\sigma=\frac{-1}{4\pi
D}\left[\sqrt{1-\frac{2M}{D}}-\sqrt{1-\frac{U(D)}{DH}}\right],\\\label{37}
&\nu=\frac{1}{8\pi
D}\left[\frac{1-\frac{M}{D}}{\sqrt{1-\frac{2M}{D}}}-\frac{1-\frac{3U(D)}{2DH}+\frac{U'(D)}{2H}}{\sqrt{1-\frac{U(D)}{DH}}}\right].
\end{align}
where
\begin{eqnarray*}
U(D)=\int[\rho_{0}(\chi_{1}-\chi_{3})+D_{0}]\frac{D^{2}}{H}dr.
\end{eqnarray*}
With the help of the definition of surface energy density, the mass of thin
shell can be found as under
\begin{equation}\label{38}
m_{s}=4\pi
D^{2}\sigma=D\left[\sqrt{1-\frac{U(D)}{DH}}-\sqrt{1-\frac{2M}{D}}\right],
\end{equation}
where
\begin{equation}\label{39}
M=\frac{U(D)}{2H}+m_{s}\sqrt{1-\frac{U(D)}{HD}}-\frac{m_{s}^{2}}{2D},
\end{equation}
describes the total amount of matter distribution within the static
irrotational spherically symmetric gravastars.

\subsection{Constant $R,~T$ and $Q$}

In this background, Eqs.\eqref{36} and \eqref{37} give
\begin{align}\label{36n}
&\sigma=\frac{-1}{4\pi
D}\left[\sqrt{1-\frac{2M}{D}}-\sqrt{1-\frac{D^{2}[\rho_{0}(\chi_{1}-\chi_{3})+D_{0}}{3H}}\right],\\\label{37n}
&\nu=\frac{1}{8\pi
D}\left[\frac{1-\frac{M}{D}}{\sqrt{1-\frac{2M}{D}}}-\frac{1-\frac{2D^{2}[\rho_{0}(\chi_{1}-\chi_{3})+D_{0}]}{3H}}
{\sqrt{1-\frac{D^{2}[\rho_{0}(\chi_{1}-\chi_{3})+D_{0}]}{3H}}}\right],
\end{align}
while an equation analogous to Eq.\eqref{38} is found to be
\begin{equation}\label{38n}
m_{s}=4\pi
D^{2}\sigma=D\left[\sqrt{1-\frac{D^{2}[\rho_{0}(\chi_{1}-\chi_{3})+D_{0}]}{3H}}-\sqrt{1-\frac{2M}{D}}\right],
\end{equation}
where
\begin{equation}\label{39n}
M=\frac{D^{3}[\rho_{0}(\chi_{1}-\chi_{3})+D_{0}]}{6H}+m_{s}\sqrt{1-\frac{D^{2}[\rho_{0}(\chi_{1}-\chi_{3})+D_{0}]}{3H}}-\frac{m_{s}^{2}}{2D},
\end{equation}
indicates the total amount of fluid configurations inside the static
non-rotating self-gravitating spherical gravastar like structures.

As described by Yousaf \emph{et al.} \cite{27}, to match the exterior and interior metrics at the hyper-surface, the following two conditions should be fulfilled for the thin shell at the boundary
\begin{align}\label{27n1}
f_{,RR}[\partial_y R|_-^+=0,\quad f_{,RR}K^*_{\lambda\sigma}|_-^+=0,\quad f_{,QQ}[\partial_y Q|_-^+=0,\quad K|_-^+=0,
\end{align}
along with
\begin{align}\label{27n2}
R|_-^+=0,\quad Q|_-^+=0,
\end{align}
where $K^*_{\lambda\sigma}$ and $K$ are the trace-free and trace components of the extrinsic curvature tensor. These equations describe the matching conditions for $f(R,T,Q)$ theory, provided
$f_{,RR}\neq0$ and $f_{,QQ}\neq0$ are fulfilled.
For the continuity of $R$ and $Q$ in an environment of thin shells, the fulfillment of Eqs.\eqref{27n1} and \eqref{27n2} at the boundary is required.

The above mentioned conditions have been evaluated by Yousaf \emph{et al.} \cite{27}. They considered the interior, exterior and the induced metrics for the cylindrically symmetric regions involving the non-dissipative anisotropic matter. Then, they developed the dynamical equations using the perturbation scheme and the contraction of Bianchi identities.

\section{Physical Features Of The Model}

This section is devoted to analyze few characteristics in order to
model a viable and well-consistent spherically symmetric isotropic
gravastar.

\subsection{Shell's Proper Length}

\begin{center}\begin{figure} \centering
\epsfig{file=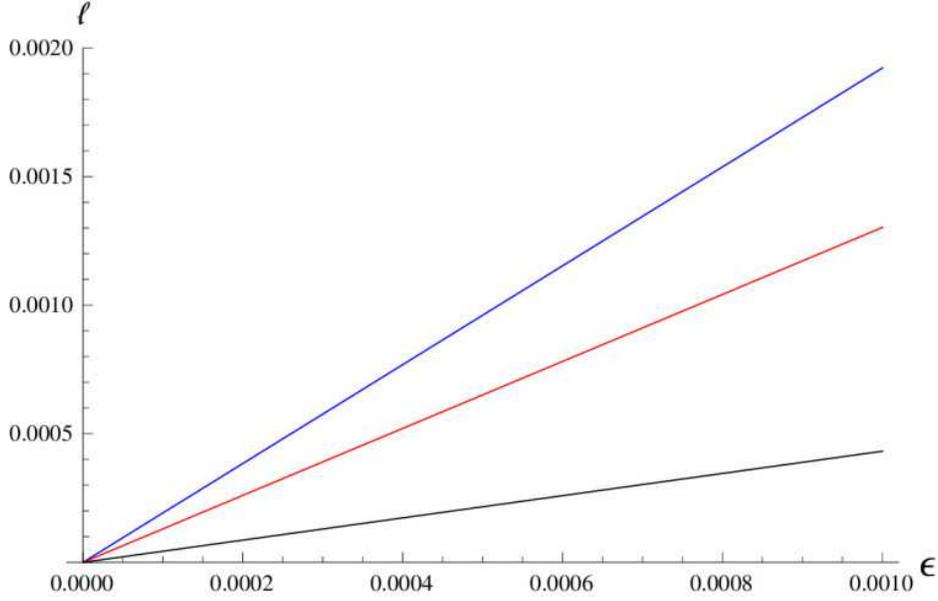,width=0.9\linewidth} \caption{Proper length
versus the thickness of the shell}\label{f2}
\end{figure}\end{center}

The proper length of the shell can be calculated from the outer
boundary of interior region $r=D$ to the outer boundary of the shell
$r=D+\epsilon$ where $\epsilon\ll1$ and indicates the small
variations. Here, we shall denote length with the letter $l$. Thus,
one can define proper length between the two surfaces as follows
\begin{equation}
l=\int_{D}^{D+\epsilon}\sqrt{e^{\lambda}}dr=\int_{D}^{D+\epsilon}\frac{dr}{\sqrt{2\ln[r]+\frac{r^{2}\psi_{1}}{2H}+C}}.
\end{equation}
As the integration of the above equation is not possible, therefore,
we shall solve this problem through numerical technique. We have
plotted a graph in order to see the physical applicability of such
results on the structure of gravastars. In Fig.\eqref{f2}, we have
seen the abrupt change in the radial profile of gravastars, which is
what we expect for the gravastars structures.

\subsection{Energy Content}

\begin{center}\begin{figure} \centering
\epsfig{file=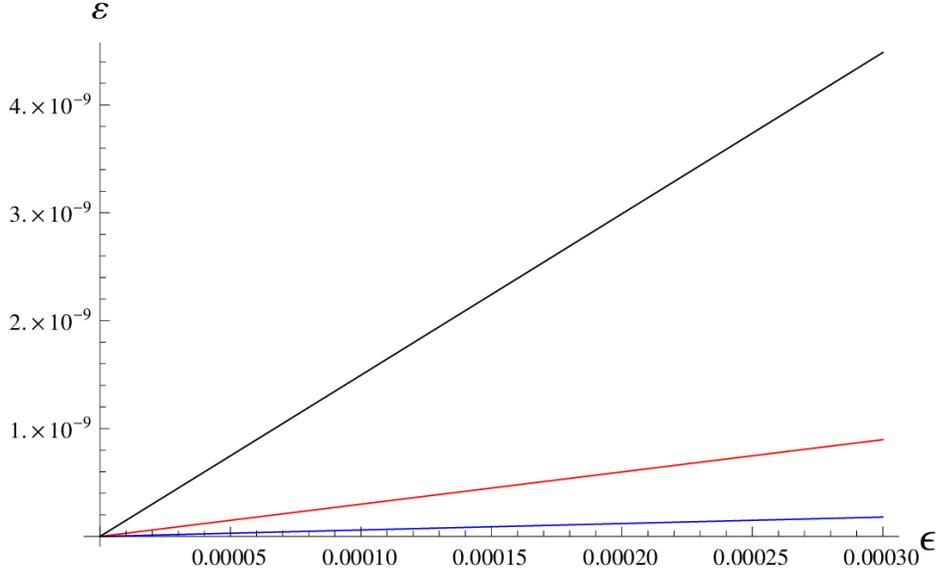,width=0.9\linewidth} \caption{Graph for
energy related with thickness of the shell} \label{f3}
\end{figure}\end{center}

Energy of the gravastar comes from the inner region having the
equation of state  $\rho=-P$ . The negative pressure
shows here the repulsive nature of the energy which desist to form
the singularity in gravastar. However, the energy within the shell
comes out to be
\begin{equation}
\varepsilon=\int_{D}^{D+\epsilon}\rho^{(eff)}r^{2}dr=\int_{D}^{D+\epsilon}\frac{Z(2+r^{2}\psi_{1})r}{\psi_{3}}dr.
\end{equation}
By considering the thin shell approximation, we have numerically
solved the above equation and draw the corresponding graph. The
Fig.\eqref{f3} indicates the direct relationship of the shell with
its thickness.

\subsection{Entropy}

\begin{center}\begin{figure} \centering
\epsfig{file=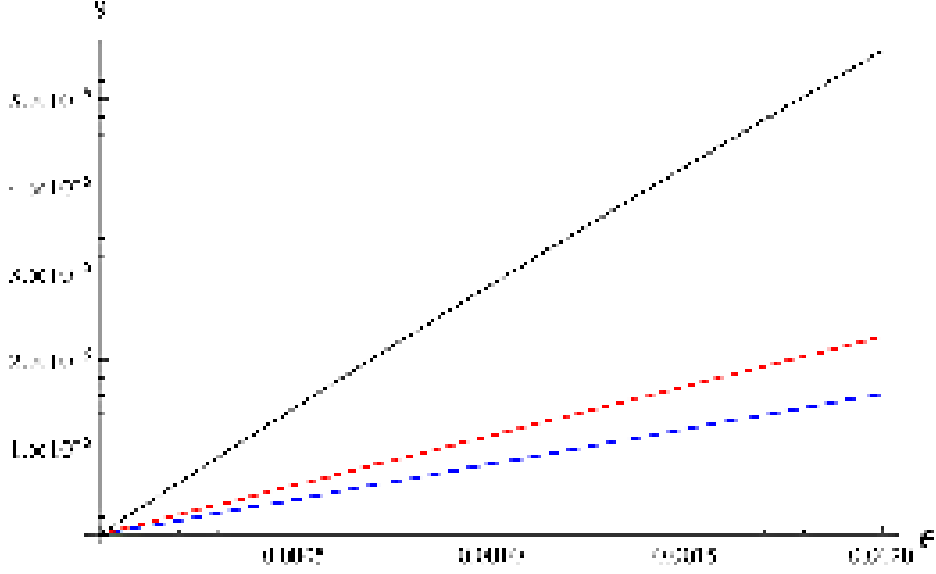,width=0.9\linewidth} \caption{Graph for
entropy related with thickness of the shell} \label{f4}
\end{figure}\end{center}

According to the model of gravastar presented by Mazur and Mottola
\cite{28}, the entropy $S$ of the shell can be
found from the following formula as
\begin{equation}\label{44}
S=\int_{D}^{D+\epsilon}s(r)r^{2}\sqrt{e^{\lambda}}dr,
\end{equation}
where $s(r)$ denotes the entropy density and can be expressed as
follows
\begin{equation}\label{45}
s(r)=\frac{\alpha^{2}k_{B}^{2}T(r)}{4\pi\hbar^{2}}=\alpha\left(\frac{k_{B}}{\hbar}\right)\sqrt{\frac{p^{(eff)}}{2\pi}}.
\end{equation}
where $\alpha$ is a dimensionless constant. In the present work, we
shall use geometric as well as Planckian units, under which one
needs to take $G=c=1$ along with $k_B=\hbar=1$. Then the entropy
density takes the form
\begin{equation}\label{46}
s(r)=\alpha\sqrt{\frac{p^{(eff)}}{2\pi}}.
\end{equation}
Then Eq.\eqref{44} takes the form
\begin{equation}\label{47}
S=\int_{D}^{D+\epsilon}\frac{\alpha}{\sqrt{2\pi}}\sqrt{\frac{Z(2+r^{2}\psi_{1})}{\psi_{3}
r}}r^{2}\frac{1}{\sqrt{2\ln[r]+\frac{r^{2}\psi_{1}}{2H}}+C}dr.
\end{equation}
We have plotted a figure \eqref{f4} in order to study the extend of
disorderness for the isotropic spherical gravastars with respect to
shell thickness. One can notice from the Fig.\eqref{f4} that the
entropy of the spherical gravastars is turned out to be zero which
the corresponding shell thickness is zero. This is one of viable
criteria for the single condensate state of celestial object as
provided by \cite{mazur2004gravitational}.

\subsection{Equation Of State}

One can define barotropic EoS at $r=D$ with respect to the effective
variables as
\begin{equation}\label{49}
\nu=\omega(D)\sigma.
\end{equation}
This equation relates the state variables which could help us to
describe the physical properties of the system. In our system the
state variables are surface energy density and surface pressure.
With the help of Eqs.\eqref{36} and \eqref{37}, the state parameter
can be found as
\begin{equation}\label{50}
\omega(D)=\frac{1}{2}\left[\frac{\frac{1-\frac{M}{D}}{\sqrt{1-\frac{2M}{D}}}
-\frac{1-\frac{3U(D)}{2HD}+\frac{U'(D)}{2H}}{\sqrt{1-\frac{U(D)}{HD}}}}{\sqrt{1-\frac{U(D)}{H}-\sqrt{1-\frac{2M}{D}}}}\right].
\end{equation}
To obtain real values of $\omega(D)$, the terms
appearing in the numerator and denominator of Eq.\eqref{50} are to be taken positive. After
applying the binomial series on Eq.\eqref{50}, we have
\begin{equation}\label{51}
\omega(D)\approx\frac{1}{2}\left[\frac{2H-HD\frac{U'(D)}{U(D)}-\frac{U'(D)}{2}}{\frac{2MH}{U(D)}-1}\right].
\end{equation}
It can be analyzed from Eq.\eqref{51}, that the constraints
$\frac{M}{U(D)}>1$ along with
$2>D\frac{U'(D)}{U(D)}+\frac{U'(D)}{2H}$ and $\frac{M}{U(D)}<1$ with
$2<D\frac{U'(D)}{U(D)}+\frac{U'(D)}{2H}$ will give positive value to
$\omega(D)$, while $\omega(D)$ will have a negative value for
$\frac{M}{U(D)}>1$ along with
$2<D\frac{U'(D)}{U(D)}+\frac{U'(D)}{2H}$ and $\frac{M}{U(D)}<1$
along with $2>D\frac{U'(D)}{U(D)}+\frac{U'(D)}{2H}$ relations. In
view of present choices of curvature variables, the state parameter
can be expressed as
\begin{equation}\label{50n}
\omega(D)=\frac{1}{2}\left[\frac{\frac{1-\frac{M}{D}}{\sqrt{1-\frac{2M}{D}}}
-\frac{1-\frac{2D^{2}[\rho_{0}(\chi_{1}-\chi_{3})+D_{0}]}{3H}}{\sqrt{1
-\frac{D^{2}[\rho_{0}(\chi_{1}-\chi_{3})+D_{0}]}{3H}}}}{\sqrt{1-\frac{D^{2}[\rho_{0}(\chi_{1}-\chi_{3})+D_{0}]}{3H}-\sqrt{1-\frac{2M}{D}}}}\right],
\end{equation}
In order to avoid complex values of $\omega(D)$, we need to take positive values of
numerator and denominator of the above expression. After the application of binomial series on the
numerator and denominator of Eq.\eqref{50n}, one can find that
\begin{equation}\label{51n}
\omega(D)\approx\frac{1}{2}\left[\frac{3}{\frac{6HM}{D^{3}[\rho_{0}(\chi_{1}-\chi_{3})+D_{0}]}-1}\right],
\end{equation}
In Eq.\eqref{51n}, the positive value of $\omega(D)$ can be achieved
by taking
$\frac{M}{D^{3}}>\frac{\rho_{0}(\chi_{1}-\chi_{3})+D_{0}}{6H}$,
while its negative can be observed on taking
$\frac{M}{D^{3}}<\frac{\rho_{0}(\chi_{1}-\chi_{3})+D_{0}}{6H}$. The
positive (for instance +1) and the negative values of $\omega(D)$
(for example -1) could provide an effective platform to model
interior and thin shell of gravastars. Such kind of situations might
be useful to understand the mathematical modeling of spherical
relativistic isotropic gravastars.

\section{Conclusion}

This paper is devoted to analyze the role of $f(R,T,Q)$ theory on
the formation of relativistic isotropic gravastars. The structures like
gravastar, has been considered to be an alternative form of the
black hole. To inspect gravastar in
this modified theory, we considered suitable static metric (which is spherically symmetric) and then determined its metric coefficients for the three regions separately.  Afterwards the physical characteristics edify the  substantial results. The outer to boundary surface has been taken to be vacuum. For this purpose, we have considered a Schwarzschild space time and match it by evaluating mathematical constraints presented by Israel and Darmois. With this background, we calculated a set of the collapsing star's exact and singularity-free models that could help to portrays physically reasonable aspects of gravastars in $f(R,T,Q)$ gravity. These are given as follows:\\
(1)\emph{Description of Pressure and Density:} We choose an arbitrary value of radius $r$ and its parameters, by doing that we deduced the effective density and effective pressure displayed persistent deviation even in the presence of $f(R,T,Q)$ theory. In the present paper, we found that at radius $r=1.0010$, the value of density is $\rho=1.2008\times10^{-6}$. This shows negligible change in the density, while Yousaf \cite{53} found that at $r=1$ the value of density is $\rho=1.32$, which indicates the small but considerable change in the density.\\
(2)\emph{Proper Length:} We have studied the behavior of the proper
length in account of shell thickness. Under the discussion of the
effective matter, Fig.\eqref{f2} indicates that the length of the
shell increases in correspondence to its thickness. This result
is also consistent with the results obtained by Yousaf \cite{53} for cylindrical gravastar structures in $f(R,T)$ gravity. In this work, the thickness of the shell is found to be $\epsilon=0.0010$ for the length $\ell=0.0018$, while Yousaf \cite{53} declared  $\epsilon$ to be $1.0$ for $0.4$ length $\ell$.\\
(3)\emph{Energy:} A graph describing the variation of the energy with respect to the shell
thickness in the presence of $f(R,T,Q)$ terms has been drawn and mentioned in Fig\eqref{f3}.
The graph confirms the direct proportionality between the energy and thickness of the shell in the presence of effective matter.
In this paper, the negligible thickness, i.e., $\epsilon=0.00030 km$ gives the small amount of energy i.e., $\varepsilon=4\times10^{-9}$. while, Yousaf \cite{53} found a large amount of energy within the shell, (i.e., $\varepsilon=250$) at the small thickness of the shell (i.e., $\epsilon=1.4$).\\
(4)\emph{Entropy:} We have plotted a graph describing a relation of
entropy of the system with the inner shell. Figure \eqref{f4}
indicates that the entropy enhances the thickness of the shell, thus
enhancing its role under the influence of effective matter. In the recent paper at thickness $\epsilon=0.0010$, we found the value of the corresponding entropy to be $S=5\times10^{-6}$. However, Yousaf \cite{53} found the entropy of the shell to be $S=0.5$ for the thickness $\epsilon=1.4$.\\
(5)\emph{Equation of State:} We have also computed the ranges of
equation of state parameter, under which it is positive or negative.
From Eq.\eqref{51n}, we have noticed that the constraint
$\frac{M}{D^{3}}>\frac{\rho_{0}(\chi_{1}-\chi_{3})+D_{0}}{6H}$ will
give us the positive value of state parameter, while its negative
value can be observed on taking
$\frac{M}{D^{3}}<\frac{\rho_{0}(\chi_{1}-\chi_{3})+D_{0}}{6H}$.

\vspace{0.5cm}

{\bf Acknowledgments}

\vspace{0.25cm}

The works of ZY and MZB were supported by National Research Project
for Universities (NRPU), Higher Education Commission, Islamabad
under the research project No. 8754/Punjab/NRPU/R\&D /HEC/2017.

\vspace{0.3cm}

\section*{Appendix}

The values of $\chi_{i}'s$ where $i=1,2,3,...,9,10$ and $D_{j}$ where $j=0,1,2$ appearing in equations \eqref{11}, \eqref{12} and \eqref{13} represent the effects of $f(R,T,Q)$ and the dark source, respectively.
\begin{align*}
\nonumber\chi_{1}&=(f_{T}+\frac{Rf_{Q}}{2}+1+f_{T}e^{-\nu}+\frac{\nu'\lambda'f_{Q}e^{-\nu}}{8}-\frac{3\nu''f_{Q}e^{-\lambda}}{4}
-\frac{3\nu'f_{Q}e^{-\lambda}}{2r}+\frac{f_{Q}''e^{-\lambda}}{2}\\\nonumber
&-\frac{3\nu'^{2}e^{-\lambda}f_{Q}}{8}-\frac{f_{Q}'\lambda'e^{-\lambda}}{4}
+\frac{e^{-\lambda}f_{Q}'}{r}+\frac{\nu'\lambda'f_{Q}e^{\nu-\lambda}}{4}),
\\\nonumber
\chi_{2}&=f_{Q}'e^{-\lambda}-\frac{f_{Q}\lambda'e^{-\lambda}}{4}+\frac{f_{Q}e^{-\lambda}}{r},
\\\nonumber
\chi_{3}&=\frac{-f_{Q}''e^{-\lambda}}{2}+\frac{f_{Q}'\lambda'e^{-\lambda}}{4}-\frac{\lambda'\nu'e^{-\nu}f_{Q}}{8}+\frac{\nu''e^{-\lambda-\nu}f_{Q}}{4}
-\frac{f_{Q}'e^{-\lambda}}{r}+\frac{\nu'e^{-\nu-\lambda}f_{Q}}{2}\\\nonumber
&+\frac{\nu'^{2}e^{-\lambda}f_{Q}}{8},
\\\nonumber
\chi_{4}&=-f_{Q}'e^{-\lambda}+\frac{f_{Q}\lambda'e^{-\lambda}}{4}-\frac{f_{Q}e^{-\lambda}}{r},
\\\nonumber
D_{0}&=\frac{f}{2}-\frac{f_{R}R}{2}+f_{R}''e^{-\lambda}-\frac{e^{-\lambda}f_{R}'\lambda'}{2}+\frac{2e^{-\lambda}f_{R}'}{r}.
\\\nonumber
\chi_{5}&=-f_{T}-\frac{e^{-\lambda}f_{Q}'\nu'}{4}-\frac{\lambda'\nu'f_{Q}e^{-\nu}}{8}+\frac{\nu''f_{Q}e^{-\lambda}}{4}+\frac{f_{Q}\nu'e^{-\lambda}}{2r}
+\frac{\nu'^{2}f_{Q}e^{-\lambda}}{8},
\\\nonumber
\chi_{6}&=f_{T}+\frac{Rf_{Q}}{2}+1+\frac{f_{Q}'\nu'e^{-\lambda}}{4}+\frac{3\lambda'\nu'f_{Q}e^{-\lambda}}{8}-\frac{3\nu''f_{Q}e^{-\lambda}}{4}
+\frac{\nu'f_{Q}e^{-\lambda}}{r}\\\nonumber
&-\frac{3\nu'^{2}e^{-\lambda}f_{Q}}{8}+\frac{2\lambda'f_{Q}e^{-\lambda}}{r},
\\\nonumber
\chi_{7}&=\frac{\lambda'f_{Q}e^{-\lambda}}{4}+\frac{\nu'f_{Q}e^{-\lambda}}{4}+\frac{2f_{Q}e^{-\lambda}}{r},
\\\nonumber
D_{1}&=\frac{-f}{2}+\frac{Rf_{R}}{2}-\frac{\nu'f_{R}'e^{-\lambda}}{2}-\frac{2f_{R}'e^{-\lambda}}{r},
\\\nonumber
\chi_{8}&=-f_{T}+\frac{f_{Q}'\nu'e^{-\lambda}}{4}-\frac{\lambda'\nu'e^{-\nu}f_{Q}}{8r^{2}}+\frac{\nu''f_{Q}e^{-\lambda}}{4}+\frac{\nu'f_{Q}e^{-\lambda}}{2r}
+\frac{\nu'^{2}e^{-\lambda}f_{Q}}{8},
\\\nonumber
\chi_{9}&=f_{T}+\frac{Rf_{Q}}{2}+1+f_{Q}''e^{-\lambda}-\frac{f_{Q}'\lambda'e^{-\lambda}}{2}+\frac{3f_{Q}'\nu'e^{-\lambda}}{4}
-\frac{f_{Q}'\nu'e^{-\lambda}}{8}+\frac{\nu''e^{-\lambda}f_{Q}}{4}\\&-\frac{f_{Q}\nu'e^{-\lambda}}{2r}+\frac{\nu'^{2}e^{^{-\lambda}f_{Q}}}{8}
+\frac{f_{Q}'e^{-\lambda}}{r}+\frac{2f_{Q}}{r^{2}}-\frac{2f_{Q}e^{-\lambda}}{r^{2}}+\frac{\lambda'f_{Q}e^{-\lambda}}{r},
\\\nonumber
\chi_{10}&=\frac{3\nu'f_{Q}e^{-\lambda}}{4}+\frac{3f_{Q}e^{-\lambda}}{r}+2f_{Q}'e^{-\lambda}-\frac{f_{Q}e^{-\lambda}\lambda'}{2},
\\\nonumber
D_{2}&=\frac{-f}{2}+\frac{f_{R}R}{2}-\frac{f_{R}'e^{-\lambda}\nu'}{2}-f_{R}''e^{-\lambda}+\frac{e^{-\lambda}f_{R}'\lambda'}{2}-\frac{e^{-\lambda}f_{R}'}{r}.
\end{align*}
This is the value of $Z$ which has been appeared in non-conserved \eqref{14} shows the effects of modified theory along with dark source effects.
\begin{align*} \nonumber
Z&=\frac{2}{2+Rf_{Q}+2f_{T}}\left[\rho'f_{Q}e^{-\lambda}(-\nu''-\frac{\nu'^{2}}{2}+\frac{\nu'\lambda'}{2}-\frac{2\nu'}{r})
-\frac{f_{Q}Pe^{-\lambda}\nu''\lambda'}{2}\right.\\\nonumber
&-\rho'f_{T}-\rho
f_{T}'-\frac{f_{T}\rho'}{2}+\frac{f_{T}P'}{2}-\frac{f_{Q}P'}{r^{2}}\left(1-e^{-\lambda}-\frac{re^{-\lambda}\nu'}{2}\right)-\frac{f_{T}e^{\nu}P'}{r^{2}}\\\nonumber
&+\frac{f_{Q}'\rho
e^{-\lambda}}{r^{2}}\left(-1+e^{\lambda}-\nu'r\right)+
\frac{e^{-\lambda}f_{Q}\rho}{r^{2}}\left(-1+e^{\lambda}-\nu'r\right)-\frac{Pf_{Q}e^{-\lambda}\lambda'\nu'^{2}}{4}\\\nonumber
&+\frac{Pf_{Q}e^{-\lambda}\lambda'^{2}\nu'}{4}+\frac{Pf_{Q}e^{-\lambda}\lambda'^{2}}{r}-\frac{f_{Q}e^{-\lambda}\nu''P'}{4}
-\frac{f_{Q}e^{-\lambda}\nu'^{2}P'}{8}\\\nonumber
&+\frac{f_{Q}e^{-\lambda}\nu'\lambda'P'}{8}+\frac{Pf_{Q}\nu''e^{-\lambda}}{r}-\frac{Pf_{Q}\nu'\lambda'e^{-\lambda}}{r}
-\frac{Pf_{Q}e^{-\lambda}\nu'}{r^{2}}\\\nonumber
&+\left.\frac{2Pf_{Q}}{r^{3}}-\frac{2Pf_{Q}e^{-\lambda}}{r^{3}}+f_{T}'P
+\frac{f_{Q}'P}{2}\left(\frac{2\nu'e^{-\lambda}}{r}-\frac{2}{r^{2}}+\frac{2e^{-\lambda}}{r^{2}}\right)\right].
\end{align*}


\vspace{0.5cm}

\begin{thebibliography}{40}


\bibitem{1} D. Pietrobon, A. Balbi, and D. Marinucci \emph{Phys. Rev. D}, vol. 74,
p. 043524, 2006.
\bibitem{2} T. Giannantonio et al. \emph{Phys. Rev. D}, vol. 74, p. 063520, 2006.
\bibitem{3} A. G. Riess et al. \emph{Astrophys. J.}, vol. 659, p. 98, 2007.
\bibitem{4} S. Nojiri and S. D. Odintsov \emph{Phys. Rev. D}, vol. 74, no. 8, p. 086005,
2006.
\bibitem{5} E. J. Copeland, M. Sami, and S. Tsujikawa \emph{Int. J. Mod. Phys. D}, vol. 15,
p. 1753, 2006.
\bibitem{6} K. Bamba, S. Capozziello, S. Nojiri, and S. D. Odintsov \emph{Astrophys.
Space Sci.}, vol. 342, p. 155, 2012.
\bibitem{7} S. Nojiri and S. D. Odintsov \emph{Phys. Rep.}, vol. 505, p. 59, 2011.
\bibitem{8} S. Nojiri, S. D. Odintsov, and V. K. Oikonomou \emph{Phys. Rep.}, vol. 692,
p. 1, 2017.
\bibitem{9} S. Capozziello and V. Faraoni, \emph{Beyond Einstein gravity: A Survey of
gravitational theories for cosmology and astrophysics}, vol. 170. Springer
Science \& Business Media, 2010.
\bibitem{10} S. Capozziello and M. De Laurentis \emph{Phys. Rep.}, vol. 509, p. 167, 2011.
\bibitem{11} A. De Felice and S. Tsujikawa \emph{Living Rev. Relativ.}, vol. 13, p. 3, 2010.
\bibitem{12} A. Joyce, B. Jain, J. Khoury, and M. Trodden \emph{Phys. Rep.}, vol. 568, p. 1,
2015.
\bibitem{13} Y.-F. Cai, S. Capozziello, M. De Laurentis, and E. N. Saridakis \emph{Rep.
Prog. Phys.}, vol. 79, p. 106901, 2016.
\bibitem{14} Z. Yousaf, K. Bamba, and M. Z. Bhatti \emph{Phys. Rev. D}, vol. 93, p. 124048,
2016.
\bibitem{15} Z. Yousaf, K. Bamba, and M. Z. Bhatti \emph{Phys. Rev. D}, vol. 93, p. 064059,
2016.
\bibitem{16} K. Bamba and S. D. Odintsov \emph{Symmetry}, vol. 7, p. 220, 2015.
\bibitem{17} M. F. Shamir and A. Malik \emph{Comm. Theor. Phys.}, vol. 71, p. 599, 2019.
\bibitem{18} T. Harko, F. S. N. Lobo, S. Nojiri, and S. D. Odintsov \emph{Phys. Rev. D},
vol. 84, p. 024020, 2011.
\bibitem{19} Z. Haghani, T. Harko, F. S. N. Lobo, H. R. Sepangi, and S. Shahidi
\emph{Phys. Rev. D}, vol. 88, p. 044023, 2013.
\bibitem{20} I. Ayuso, J. B. Jim\'{e}nez, and \'{A}. de la Cruz-Dombriz \emph{Phys. Rev. D},
vol. 91, p. 104003, 2015.
\bibitem{21} S. D. Odintsov and D. S\'{a}ez-G\'{o}mez \emph{Phys. Lett B}, vol. 725, p. 437, 2013.
\bibitem{22} E. H. Baffou, M. J. S. Houndjo, and J. Tosssa \emph{Astrophys. Space Sci.},
vol. 361, p. 376, 2016.
\bibitem{23} G. Dvali New \emph{J. Phys.}, vol. 8, p. 326, 2006.
\bibitem{24} Z. Yousaf, M. Z. Bhatti, and U. Farwa \emph{Class. Quantum Grav.}, vol. 34,
p. 145002, 2017.
\bibitem{25} Z. Yousaf, M. Z. Bhatti, and U. Farwa \emph{Mon. Not. Roy. Astron. Soc.},
vol. 464, p. 4509, 2016.
\bibitem{26} Z. Yousaf, K. Bamba, M. Z. Bhatti, and U. Farwa \emph{Eur. Phys. J. A},
vol. 54, p. 122, 2018.
\bibitem{27} Z. Yousaf, M. Z. Bhatti, and U. Farwa \emph{Eur. Phys. J. C}, vol. 77, p. 359,
2017.
\bibitem{28} P. O. Mazur and E. Mottola \emph{Proc. Natl. Acad. Sci.} U.S.A, vol. 101,
p. 9545, 2004.
\bibitem{29} K. S. Virbhadra, D. Narasimha, and S. M. Chitre \emph{Astron. Astrophys.},
vol. 337, p. 1, 1998.
\bibitem{30} K. S. Virbhadra and G. F. R. Ellis \emph{Phys. Rev. D}, vol. 65, p. 103004,
2002.
\bibitem{31} M. Z. Bhatti and Z. Yousaf \emph{Int. J. Mod. Phys. D}, vol. 26, p. 1750045,
2017.
\bibitem{32} Z. Yousaf, M. Z. Bhatti, and M. F. Malik \emph{Eur. Phys. J. Plus}, vol. 134,
p. 470, 2019.
\bibitem{33} Z. Yousaf, M. Z. Bhatti, and S. Yaseen \emph{Eur. Phys. J. Plus}, vol. 134,
p. 487, 2019.
\bibitem{34} M. Z. Bhatti, Z. Yousaf, and M. Yousaf \emph{Phys. Dark Universe}, vol. 28,
p. 100501, 2020.
\bibitem{35} M. Visser and D. L. Wiltshire \emph{Class. Quantum Grav.}, vol. 21, no. 4,
p. 1135, 2004.
\bibitem{36} C. Cattoen, T. Faber, and M. Visser \emph{Class. Quantum Grav.}, vol. 22,
no. 20, p. 4189, 2005.
\bibitem{37} C. B. M. H. Chirenti and L. Rezzolla \emph{Class. Quantum Grav.}, vol. 24,
p. 4191, 2007.
\bibitem{38} N. Bili\'{c}, G. B. Tupper, and R. D. Viollier \emph{J. Cosmol. Astropart. Phys.},
vol. 2006, p. 013, 2006.
\bibitem{39} D. Horvat, S. Iliji\'{c}, and A. Marunovi\'{c} \emph{Class. Quantum Grav.}, vol. 28,
p. 195008, 2011.
\bibitem{40} N. Sakai, H. Saida, and T. Tamaki \emph{Phys. Rev. D}, vol. 90, p. 104013,
2014.
\bibitem{41} C. B. Chirenti and L. Rezzolla \emph{Phys. Rev. D}, vol. 78, p. 084011, 2008.
\bibitem{42} F. Rahaman, A. A. Usmani, S. Ray, and S. Islam \emph{Phys. Lett. B}, vol. 717,
p. 1, 2012.
\bibitem{43} P. Pani, E. Berti, V. Cardoso, Y. Chen, and R. Norte \emph{Phys. Rev. D},
vol. 80, p. 124047, 2009.
\bibitem{44} S. Ghosh, F. Rahaman, B. Guha, and S. Ray \emph{Phys. Lett. B}, vol. 767,
p. 380, 2017.
\bibitem{45} A. Das, S. Ghosh, B. Guha, S. Das, F. Rahaman, and S. Ray \emph{Phys. Rev.
D}, vol. 95, p. 124011, 2017.
\bibitem{46} M. F. Shamir and M. Ahmad \emph{Phys. Rev. D}, vol. 97, p. 104031, 2018.
\bibitem{47} Z. Yousaf \emph{Astrophys. Space Sci.}, vol. 363, p. 226, 2018.
\bibitem{48} Z. Yousaf \emph{Eur. Phys. J. Plus}, vol. 134, p. 245, 2019.
\bibitem{49} Z. Yousaf, K. Bamba, M. Z. Bhatti, and U. Ghafoor \emph{Phys. Rev. D},
vol. 100, p. 024062, 2019.
\bibitem{50} M. Z. Bhatti, Z. Yousaf, and M. Ajmal \emph{Int. J. Mod. Phys. D}, vol. 28,
p. 1950123, 2019.
\bibitem{51} M. Z. Bhatti \emph{Mod. Phys. Lett. A}, vol. 34, p. 2050069, 2020.
\bibitem{52} M. Sharif and A. Waseem \emph{Astrophys. Space Sci.}, vol. 364, p. 189, 2019.
\bibitem{53} Z. Yousaf \emph{Phys. Dark Universe}, vol. 28, p. 100509, 2020.
\bibitem{54} Y. B. Zeldovich \emph{Mon. Not. Roy. Astron. Soc.}, vol. 160, p. 1P, 1972.
\bibitem{55} M. S. Madsen, J. P. Mimoso, J. A. Butcher, and G. F. R. Ellis \emph{Phys.
Rev. D}, vol. 46, p. 1399, 1992.
\bibitem{56} P. S. Wesson \emph{Vistas Astron.}, vol. 29, p. 281, 1986.
\bibitem{57} T. M. Braje and R. W. Romani \emph{Astrophys. J.}, vol. 580, p. 1043, 2002.
\bibitem{58} L. P. Linares, M. Malheiro, and S. Ray \emph{Int. J. Mod. Phys. D}, vol. 13,
p. 1355, 2004.
\bibitem{59} W. Israel \emph{Nuovo Cimento B}, vol. 44, no. 1, 1966.
\bibitem{60} G. Darmois \emph{Gauthier-Villars, Paris}, vol. 25, 1927.
\bibitem{61} K. Lanczos \emph{Ann. Phys.(Berl.)}, vol. 379, p. 518, 1924.
\bibitem{62} N. Sen \emph{Ann. Phys.(Berl.)}, vol. 378, p. 365, 1924.
\bibitem{63} G. Perry and R. B. Mann \emph{Gen. Relativ. Gravit.}, vol. 24, p. 305, 1992.
\bibitem{64} P. Musgrave and K. Lake \emph{Class. Quantum Grav.}, vol. 13, p. 1885, 1996.

\end{thebibliography}
\end{document}